\documentclass[aps,prl,reprint,superscriptaddress]{revtex4-1}
\usepackage{amsmath}
\usepackage{amssymb}
\usepackage{graphicx}
\usepackage{subfigure}
\usepackage{times}

\begin{document}

%\preprint{}

\title{Integer Discontinuity of Density Functional Theory}


\author{Mart\'in A. Mosquera}
\affiliation{Department of Chemistry, Purdue University, West Lafayette, IN 47907, USA}
\author{Adam Wasserman}
\affiliation{Department of Chemistry, Purdue University, West Lafayette, IN 47907, USA}
\affiliation{Department of Physics, Purdue University, West Lafayette, IN 47907, USA}
\email[]{awasser@purdue.edu}
%\thanks{}

% Equation environments
\def\bea{\begin{eqnarray}}
\def\eea{\end{eqnarray}}
\def\ben{\begin{equation}}
\def\een{\end{equation}}
\def\benu{\begin{enumerate}}
\def\enu{\end{enumerate}}

% density
\def\n{n}
\def\lsim {\ifmmode {\buildrel<\over\sim}}

% Scriptstyle
\def\sss{\scriptscriptstyle\rm}

% gamma subscript for scaling
\def\g{_\gamma}

% lambda superscript for coupling constant
% \def\l{^\lambda}
\def\t{\beta}
\def\lumo{_{\sss L}}
\def\homo{_{\sss H}}
\def\lfc{^{\lambda=1}}
\def\lo{^{\lambda=0}}
\def\app{^{\sss app}}
\def\XC{_{\sss XC}}
\def\HXC{_{\sss HXC}}
\def\H{_{\sss H}}
\def\X{_{\sss X}}
\def\ks{_{\sss s}}

%Martin's definitions
\newcommand{\mr}[1]{\mathrm{#1}}%I don't recommend \rm; \mr is more practical. 
\newcommand{\mc}[1]{\mathcal{#1}}
\newcommand{\mb}[1]{\mathbf{#1}}
\newcommand{\ud}{\text{d}}
\newcommand{\intdr}{\int \ud^3\br~}
\newcommand{\h}{_{\mr{h}}}
\newcommand{\Ha}{_{\mr{H}}}
\newcommand{\s}{_{\mr{s}}}
\newcommand{\dernr}[1]{\frac{\delta {#1} }{\delta n(\br)}}
\newcommand{\dernrs}[1]{\frac{\delta {#1}}{\delta n_{\sigma}(\mb{r})}}
\newcommand{\derN}[1]{\frac{\partial {#1}}{\partial N}}
\newcommand{\dss}{\displaystyle}
\newcommand{\nv}[1]{[n_{v,#1}]}
\newcommand{\snv}[1]{n_{v,#1}}
\newcommand{\KS}{_\mr{KS}}
\newcommand{\LDA}{^\mr{LDA}}
\newcommand{\ELDA}{^\mr{ELDA}}
\def\marnote#1{\marginpar{\tiny #1}}
\def\rsav{\langle r_s \rangle}
\def\invdif{\frac{1}{|\br_1 - \br_2|}}

%operators
\def\hatT{{\hat T}}
\def\hatV{{\hat V}}
\def\hatH{{\hat H}}
\def\1var{(\bx_1...\bx\N)}

% Fractions
\def\half{\frac{1}{2}}
\def\quart{\frac{1}{4}}

% Bold-face symbols
\def\bp{{\bf p}}
\def\br{{\bf r}}
\def\bR{{\bf R}}
\def\bu{{\bf u}}
\def\b1{{\bf 1}}
\def\bx{{x}}
\def\by{{y}}
\def\ba{{\bf a}}
\def\bq{{\bf q}}
\def\bj{{\bf j}}
\def\bX{{\bf X}}
\def\bF{{\bf F}}
\def\bchi{{\bf \chi}}
\def\bof{{\bf f}}

% script symbols
\def\cA{{\cal A}}
\def\cB{{\cal B}}
\def\cH{{\cal H}}

% Standard subscripts
\def\x{_{\sss X}}
\def\xc{_{\sss XC}}
\def\Hx{_{\sss HX}}
\def\Hxc{_{\sss HXC}}
\def\xj{_{{\sss X},j}}
\def\xcj{_{{\sss XC},j}}
\def\N{_{\sss N}}
\def\H{_{\sss H}}
\def\sH{^{\sss H}}
% Word sub and superscripts
\def\ext{_{\rm ext}}
\def\pot{^{\rm pot}}
\def\hyb{^{\rm hyb}}
\def\hah{^{1/2\& 1/2}}
\def\LSD{^{\rm LSD}}
\def\TF{^{\rm TF}}
\def\LDA{^{\rm LDA}}
\def\GEA{^{\rm GEA}}
\def\GGA{^{\rm GGA}}
\def\SPL{^{\rm SPL}}
\def\sce{^{\rm SCE}}
\def\PBE{^{\rm PBE}}
\def\DFA{^{\rm DFA}}
\def\VW{^{\rm VW}}
\def\helm{^{\rm unamb}}
\def\una{^{\rm unamb}}
\def\ion{^{\rm ion}}
\def\gs{^{\rm gs}}
\def\dyn{^{\rm dyn}}
\def\adia{^{\rm adia}}
\def\I{^{\rm I}}
\def\pot{^{\rm pot}}
\def\sav{^{\rm sph. av.}}
\def\unif{^{\rm unif}}
\def\LSD{^{\rm LSD}}
\def\ee{_{\rm ee}}
\def\vir{^{\rm vir}}
\def\ALDA{^{\rm ALDA}}
\def\PGG{^{\rm PGG}}
\def\GK{^{\rm GK}}

% spin indices
%\def\up{_\alpha}
%\def\dn{_\beta}
\def\up{_\uparrow}
\def\dn{_\downarrow}
\def\upp{\uparrow}
\def\dnn{\downarrow}

% Words
\def\td{time-dependent~}
\def\KS{Kohn-Sham~}
\def\DFT{density functional theory~}

%integrals
\def\fourint{ \int_{t_0}^{t_1} \! dt \int \! d^3r\ }
\def\fourintp{ \int_{t_0}^{t_1} \! dt' \int \! d^3r'\ }
\def\intx{\int\!d^4x}
\def\sph_int{ {\int d^3 r}}
\def\radint{ \int_0^\infty dr\ 4\pi r^2\ }

%journals
\def\PRA{Phys. Rev. A\ }
\def\PRB{Phys. Rev. B\ }
\def\PRL{Phys. Rev. Letts.\ }
\def\JCP{J. Chem. Phys.\ }
\def\JPCA{J. Phys. Chem. A\ }
\def\IJQC{Int. J. Quant. Chem.\ }

\def\la{{\langle\, }}
\def\ra{{\,\rangle }}
\def\infintw{ \int_{-\infty}^\infty }
\def\infint{ \int_{-\infty}^\infty dx\,}
\def\infintp{ \int_{-\infty}^\infty dx'\,}
\def\infintd3r{ \int_{-\infty}^\infty d^3r\,}
\def\intd3r{ \int d^3r\,}
\def\kinop{- \half \frac{d^2}{dx^2}}
\def\laplace1d{\frac{d^2}{dx^2}}
\def\plaplace1d{\frac{d^2}{d{x'}^2}}
\def\pdr{\frac{\partial}{\partial r}}
\def\padr2{\frac{\partial^2}{\partial r^2}}
\def\nup{n_\uparrow}
\def\ndown{n_\downarrow}
\def\p{\,|\,}
\def\lin{^{(1)}}
\def\bmu{{\vec \mu}}
\def\SMA{^{\rm SMA}}
\def\SPA{^{\rm SPA}}

\def\sF{\boldsymbol{\Phi}}
\def\sP{\boldsymbol{\cal P}}
\def\sV{{\cal V}}
\def\sN{\boldsymbol{\cal N}}
\def\st{\boldsymbol{\tau}}
\def\O{{\cal O}}
\def\F{{\cal F}}
\def\N{{\cal N}}
\def\a{{\alpha}}
\def\b{{\beta}}
\def\P{{\cal P}}
\def\hP{\hat{P}}
\def\tP{\tilde{P}}
\def\E{{\cal E}}
\def\G{{\cal G}}
\def\S{{\cal S}}

\def\bS{{\overline S}}


\date{\today}

\begin{abstract}
Density functional approximations to the exchange-correlation energy of Kohn-Sham theory, 
such as the local density approximation and generalized gradient approximations, 
lack the well-known integer discontinuity, a feature that is critical 
to describe molecular dissociation correctly. 
Moreover, standard approximations to the exchange-correlation energy 
also fail to yield the correct linear dependence of the ground-state energy 
on the number of electrons when 
this is a non-integer number obtained from the grand canonical ensemble statistics. 
We present a formal framework to restore the integer
discontinuity of any density functional approximation.
% of discrete-electron  
%states like the local density approximation, generalized gradient approximations, 
%hybrid functionals, etc. 
Our formalism derives from a formula
for the exact energy functional and a new constrained search functional 
that recovers the linear dependence of the energy on the number of electrons. 
\end{abstract}

\maketitle

%Density functional theory (DFT) is a reformulation of quantum mechanics that, 
%instead of looking for solutions of the time-independent Schr\"odinger equation,
%introduces an energy functional who is a functional of the electronic density 
%of the molecule or atom. Within the Kohn-Sham (KS) formulation of DFT, a simple approximation 
%to the exchange-correlation functional suffices to solve computationally cheaper
%single-particle Schr\"odinger equations that yield the electronic density.  
%When the latter is input into the KS energy functional, one obtains
% the ground-state energy of the molecule. 

Density Functional Theory (DFT) \cite{HK64, KS65} 
 is a useful formulation of ground-state quantum mechanics 
that offers a simple approach to estimate the electronic properties of molecules and solids \cite{B12}.
\citet{PPLB82} (PPLB) considered Mermin's extension \cite{M65} of DFT to systems that
adiabatically exchange electrons with a distant reservoir at zero-temperature. 
In this framework, the energy as a function of the electron-number 
is a series of straight lines interpolating the energies corresponding
to those of closed systems with integer numbers of electrons. PPLB found 
that the exchange-correlation (XC) potential displays a derivative discontinuity (DD) that, when 
added to the KS band gap, yields the fundamental band gap of the system (also see Ref. \cite{BGM13}).  
The DD is present in molecular dissociation: When 
two atoms are separated far apart they take on integer numbers of electrons 
to neutralize their charges, and the total energy of the system, which is nearly additive, tends to display
a DD with respect to a change in the number of electrons when 
one atom transfers its electron to the other. 

The DD of the XC energy functional and the linear dependency 
between discrete intervals is required to improve the physics of density functionals. 
The missing integer discontinuity causes problems 
in the estimation of ground-state properties like binding energies \cite{CMY08} and reaction barriers \cite{ZY98}. 
In time dependent density functional theory, the missing integer discontinuity is also 
required to improve the accuracy of density-functional approximations (DFA's) \cite{T03,FRM11,HG12,MK05,KSKV10,KS13}, especially to
describe bond-stretching processes.
%In general, most approximations in DFT are unable to describe bond stretching without 
%recurring to spin-polarized DFT, in which the symmetry is broken. 
%Another error
%that occurs at stretching chains of atoms an imposing symmetry is 
A strong delocalization 
error \cite{MCY08} occurs due to the lack of piecewise linear dependency of the resulting fragment energies
with the number of electrons. This non-linearity is pervasive and affects all calculations that use continuous XC energy functionals such as the local density approximation (LDA) \cite{CMY12}. 
These known problems point to the need to develop new functionals with the correct piecewise linearity, capable 
of describing bond-stretching without resorting to symmetry breaking.
Long-range and non-local corrections are usually added to the XC energy functional \cite{CMY07} to solve these problems. 
In most cases, the corrections improve the results without completely recovering the linear behavior of the XC energy
between integers and its DD. 

Non-empirical functionals such as the LDA and 
generalized gradient approximations work well for atoms with integer numbers of electrons.
\citet{KK13} explored the properties of a simple ensemble average of XC
energies of pure states. They showed that the piecewise linearity is almost 
restored by their approach using the optimized effective potential method. Their results illustrate
the plausibility of recovering the integer discontinuity of most functionals of 
discrete-electron states that are apparently continuous in terms of the density.    

In this work we propose a formalism to restore \emph{completely} the linear dependency on the 
electron-number between integers. We use the 
fact that most density functional approximations have been developed for 
closed systems with integer numbers of electrons. We perform 
an expansion of the ensemble XC energy functional in terms of XC and KS kinetic energies evaluated at closed,
fully interacting discrete-electron densities that sum to the correct ensemble ground-state density. We then connect 
the resulting expression to an expansion of the
KS kinetic energy evaluated at non-interacting discrete-electron densities 
that yield the same ensemble ground-state density. 
For density-functional approximations, 
a constrained search is proposed to replace the Levy-Lieb search that 
requires the electron-electron repulsion operator. This search assumes
non-interacting $v$-representability of the discrete-state densities and permits 
to recover strictly the piecewise-defined linearity for approximate XC energy functionals
and their concomitant integer discontinuities. 

The PPLB density functional is defined as:
\begin{equation}
E_v[n]=F[n]+\int d\mb{r}~ n(\mb{r})v(\mb{r})~,
\end{equation}
where $F[n]$ is the constrained-search functional:
\begin{equation}\label{levy}
F[n]=\inf_{\hat{D}\rightarrow n}\mr{Tr}\{(\hat{T}+\hat{W})\hat{D}\}~.
\end{equation}
$\hat{T}$ is the kinetic energy operator, $\hat{W}$ is the electron-electron
repulsion operator, and $\hat{D}$ is the density matrix operator in Fock space. The notation
``${\hat{D}\rightarrow n}$" indicates that the search for the infimum is performed over all density matrices
satisfying $\mr{Tr}\{\hat{D}\hat{n}(\mb{r})\}=n(\mb{r})$. In order to carry out an equivalent search without requiring this density constraint, 
we introduce the Lagrange multiplier $u[n]$ as indicated below. The generalized energy $\mc{E}_N[u]$, 
now a functional of $u[n]$, involves a search over all density matrices
corresponding to $N$ electrons ($N$ is in general non-integer):
\begin{equation}\label{thrower}
\mc{E}_N[u]=\inf_{\hat{D}\rightarrow N} \mr{Tr}\{(\hat{T}+\hat{W}+\int d\mb{r}~u(\mb{r})
\hat{n}(\mb{r}))\hat{D}\}~.
\end{equation}
Here, $N$ is a real number between $J$ and $J+1$, where $J$ is a positive integer. 
If the convexity assumption holds, i.e., $\mc{E}_{J-1}[u]-\mc{E}_{J}[u]\ge 
\mc{E}_{J}[u]-\mc{E}_{J+1}[u]$ for any $J$, then
$
\mc{E}_N[u]=(1-\omega)\mc{E}_J[u]+\omega \mc{E}_{J+1}[u],
$ 
where $\omega[n]=\int d\mb{r}~n(\mb{r})-J$. We assume that $-1<\omega<1$. 
The search for the infimum in Eq. (\ref{thrower}) yields a density matrix $\hat{D}[n]$
that is also a linear interpolation of integer-number density matrices, $\hat{D}_J$ and $\hat{D}_{J+1}$.
For example, if the bordering systems are pure ground states then $\hat{D}_M=|\psi_M\rangle \langle \psi_M|$,
$M=J,J+1$. The densities of the pure states, that is 
$n_M[u](\mb{r})=\mr{Tr}\{\hat{D}_M[u]\hat{n}(\mb{r})\}$, $M=J,J+1$, satisfy the restriction:
\begin{equation}\label{dens}
n(\mb{r})=(1-\omega)n_{J}[u](\mb{r})+\omega n_{J+1}[u](\mb{r})~.
\end{equation}
Because $u$ is a functional of the density, so are the densities $n_J$ and 
$n_{J+1}$. Inserting the minimizing density matrix $\hat{D}[n]$ into $F[n]$ we find that
\begin{equation}\label{Faverage}
F[n]=(1-\omega[n])F[n_J]+\omega[n] F[n_{J+1}]
\end{equation} 
(Note: If $-1<\omega<0$ then we replace $\omega$ by $-\omega$, and $J+1$ by $J-1$ in the above equation). 

For notational convenience, we introduce the average function:
\begin{equation}
y(x)=\begin{cases}
1&x=0~,\\
1-x& 0<x<1~,\\
1+x& -1<x<0~,\\
0&~ \mr{otherwise}~,
\end{cases}
\end{equation}
which allows us to express $F$ (as well as the energy, density, etc.) as:
\begin{equation}\label{Fensemble}
F[n]=\sum_M y(N-M)F[n_M]~,
\end{equation}
where $N=\smallint n$ is of course a density-functional as well.
The functional $F[n]$ is split in the usual Kohn-Sham manner:
\begin{equation}\label{ksplit}
F[n]=T\ks[n]+E\HXC[n]~,
\end{equation}
where $T\ks[n]=\inf\{\mr{Tr}[\hat{T}\hat{D}\ks]|\hat{D}\ks\rightarrow n\}$,
and $E\HXC[n]=E\H[n]+E\XC[n]$, the Hartree and exchange-correlation energy functionals.

The ground-state energy for the auxiliary system of non-interacting electrons, $\mc{E}_{{\sss s},N}$ 
can be thought of as a functional of $u\ks(\br)$, an analog of $u(\br)$
introduced to carry out the non-interacting search version of Eq. (\ref{thrower}):
\begin{equation}
\mc{E}_{{\sss s},N}[u\ks]=\inf_{\hat{D}\ks\rightarrow N}\mr{Tr}\{(\hat{T}+\int d\mb{r}~ u\ks(\mb{r})
\hat{n}(\mb{r}))\hat{D}\ks\}~.
\end{equation} 
As in the case of $F[n]$, $T\ks[n]$ returns two densities $n_{{\sss s},J}(\br)$ and $n_{{\sss s},J+1}(\br)$
that, when added together with the weight factor $y(N-M)$, yield the 
density $n(\br)$ of the interacting system. In what follows, we will refer to $n_{{\sss s},J}(\br)$ and 
$n_{{\sss s},J+1}(\br)$ as the non-interacting bordering-integer densities. We emphasize that even 
employing the {\em exact} exchange-correlation functional, the non-interacting integer density $n_{{\sss s},M}(\br)$ 
is {\em not} equal to the $M$-electron density of the interacting system (see Fig. 2.a
for a model system we  describe later on). Rather than being the ground-state density of $M$ 
interacting electrons in $v(\br)$ (or $M$ non-interacting electrons in $v\ks(\br)$), it is 
the ground-state density of $M$ non-interacting electrons in $u\ks(\br)$, a potential that 
differs from $v\ks(\br)$ for non-integer $M$, as illustrated in Fig. 2.b.
For example, 
$n_{{\sss s},J}(\br)=\sum_{i=1}^{J}|\phi_i(\br)|^2$, and $n_{{\sss s},J+1}(\br)=n_{{\sss s},J}(\br)+|\phi_{J+1}(\br)|^2$,
where $\{\phi_i\}(\br)$ are single-particle orbitals that satisfy 
\begin{equation}
\Big(\hat{T}+\int d\mb{r}~u\ks(\mb{r})\hat{n}(\mb{r})\Big)|\phi_i\rangle =\epsilon_i|\phi_i\rangle~,
\end{equation}
and by definition $\sum_M y(N-M)n_{{\sss s},M}(\br)=n(\br)$. For example, if $J<N<J+1$, then 
using $y$ we get that $n(\br)=n_{{\sss s},J}+\omega |\phi_{J+1}(\br)|^2$. The non-interacting bordering-integer
densities $n_{{\sss s},J}(\br)$ and $n_{{\sss s},J+1}(\br)$ are density functionals as well.
Inserting Eq. (\ref{ksplit}) on both sides of Eq. (\ref{Fensemble}) and expanding $T\ks[n]$ as 
$\sum_M y(N-M)T\ks[n_{{\sss s},M}]$, we obtain the most important result of this paper: 
\ben\label{mformula}
\begin{split}
E\HXC[n]=\sum_M y(N-&M)\Big\{ (T_{{\sss s}}[n_M]-T_{{\sss s}}[n_{{\sss s},M}])\\&+E\HXC[n_M]\Big\}~,
\end{split}
\een
an exact relation for $E\HXC[n]$ in terms of quantities that describe pure quantum states, with $T\ks$ evaluated at both,
the interacting and non-interacting bordering-integer densities. 
Eq. (\ref{mformula}) is trivially true when $n(\br)$ integrates to an integer number, but it is a 
useful identity when $J<N<J+1$ in the context of {\em approximate} DFT, as we show next.

Let us denote as $E\app\HXC[n_M]$ an approximation for $M=1,2,\ldots$ Inserting this functional
into Eq. (\ref{mformula}) yields $E\app\HXC[n]$, a useful approximation to the ensemble functional.
The densities $\{n_M\}$ can in principle be obtained from the search in $F[n]$, a functional 
we do not know. But we can circumvent the use of $F[n]$ by defining
\ben\label{gsform}
G\ks[n]=\inf_{\{\tilde{n}_M\}\rightarrow n} \sum_M y(N-M) G\ks[\tilde{n}_M]~, 
\een
where 
\ben
G\ks[\tilde{n}_M]=\inf_{\hat{D}\ks\rightarrow \tilde{n}_M} 
\mr{Tr}\{(\hat{T}+\int d\mb{r}~ v\app\HXC[\tilde{n}_M](\mb{r})\hat{n}(\mb{r}))\hat{D}\ks\}~.
\een
By $\{\tilde{n}_M\}\rightarrow n$ we refer to the constraint $\sum_M y(N-M)\tilde{n}_M(\br)=n(\br)$.
If $J<N<J+1$, the optimal densities $\{n_M\}$ that minimize the right hand side of Eq. (\ref{gsform}) are 
obtained from solving two sets of KS equations self-consistently: one with KS potential $\tilde{v}\ks[\tilde{n}_J]=v\app\HXC[\tilde{n}_J]+\tilde{u}$ and another
with $\tilde{v}\ks[\tilde{n}_{J+1}]=v\app\HXC[\tilde{n}_{J+1}]+\tilde{u}$. The orbitals arising from the KS equations
with $\tilde{v}\ks[\tilde{n}_J]$ and $\tilde{v}\ks[\tilde{n}_{J+1}]$ are complex-squared and added together to yield the 
densities $\tilde{n}_J$ and $\tilde{n}_{J+1}$. 
The external potential $\tilde{u}$ is a
Lagrange multiplier arising from the constraint $\{\tilde{n}_M\}\rightarrow n$ and 
is to be varied until the constraint is satisfied. If 
$\tilde{u}$ is set as the external potential of the system, $v$, then one obtains an approximation 
to the ensemble ground-state density. The functional in Eq. (\ref{gsform}) reformulates 
the non-interacting $v$-representability problem for an approximate XC potential. 
When the {\em exact} XC potential is used, then setting $\tilde{u}=v$ and solving 
the two sets of KS equations produces the orbitals needed to build the exact ground-state densities
$n_J$ and $n_{J+1}$.   

The total energy of the system is 
\ben
\begin{split}
E\app_v[n]=\sum_M y(N-&M)\Big(T\ks[n_M]+E\app\HXC[n_M]\\&+\int d\mb{r}~ v(\mb{r})n_M(\mb{r})\Big)~.
\end{split}
\een
The approximated ground state energy is found by setting $E\app_N[v]=\inf_{n\rightarrow N} E_v\app[n]$.
If the convexity assumption holds for our system of interest then
\ben\label{avenr}
E\app_N[v]=\sum_M y(N-M)E_M\app[v]~,
\een
where 
\ben 
E\app_M[v]=\inf_{n_M} T_{{\sss s},M}[n_M]+E\app_{{\sss HXC},M}[n_M]+\int d\mb{r} v(\mb{r})n_M(\mb{r})~.
\een
Eq. (\ref{avenr}) shows that it is possible to recover the piecewise linear dependence of 
the approximated energy. 
Using the analog of Eq. (\ref{dens}) for $J-1<N<J$ and the KS equations, it can be shown that:
\ben
\begin{split}
\dernr{E\app\HXC}=&E_J\app-E\app_{J-1}-\epsilon\app_J+v\app\HXC(\mb{r})\\
& +\sum_M y(N-M) \int d\mb{r}'~\Big(\frac{\delta E\app_{v,M}}{\delta n_M(\mb{r}')}\Big)
\frac{\delta n_M(\mb{r}')}{\delta n(\mb{r})}~.
\end{split}
\een
The term $\delta E\app_{v,M}/\delta n_M(\mb{r}')$ is a constant at the minimum 
and $\int d\mb{r}'~ \delta n_M(\mb{r}')/\delta n(\mb{r})=0$,
which leads to (dropping the Hartree contribution):
\ben
\label{dExcdn}
\dernr{E\app\XC}=-I\app-\epsilon\app_J+v\app\XC(\mb{r})~.
\een
Since $v\app\XC=\delta E\XC\app/\delta n$, by definition, we obtain the Janak's theorem \cite{J78} 
$\epsilon\app_J=-I\app$, where $I\app=E\app_{J-1}[u]-E\app_{J}[u]$ is the ionization energy 
of the system, and $J-1<N<J$. We can also write Eq. (\ref{dExcdn}) as:
\ben
\dernr{E\app\XC}
=-I\app-\dernr{T\s}-u(\mb{r})~.
\een
This result allows us to calculate the XC DD as\footnote{This discontinuity is taken along a path of ground-state
ensemble densities \cite{MW14}.}:
\ben\label{dd}
\begin{split}
\Delta\XC&=\lim_{\Delta N\rightarrow 0^+}\frac{\delta E\app\XC}{\delta n(\mb{r})}\Bigg|_{J+\Delta N}-\frac{\delta E\app\XC}{\delta n(\mb{r})}\Bigg|_{J-\Delta N}\\
&=I\app-A\app-(\epsilon\app\lumo-\epsilon\app\homo)~,
\end{split}
\een
where $A\app=E\app_{J}[u]-E\app_{J+1}[u]$ is the electron affinity of the $J$-electron system and $\epsilon\app\homo$ and $\epsilon\app\lumo$ are the HOMO and LUMO orbital energies of the $J$-electron system. 
The XC DD turns out to be the difference between the fundamental gap of the real system and the KS gap. 
However, the approximated XC DD serves the same purpose: correct the KS particle band gap. 

For an ensemble DFA the Janak's theorem is valid but the ionization theorem is not, in general. For example, 
for a system with {\slshape strictly $J$} electrons it is known that the LDA HOMO energy does not match 
the ionization predicted by LDA, i.e., when $N=J$, $\epsilon_J^{\mr{LDA}}\neq-I\LDA$. 
To satisfy the Janak's theorem, a constant must be added to the approximate XC potential to replace the 
HOMO orbital by the DFA ionization. When $J-1<N<J$, this constant is $-I-\epsilon\homo(N=J)$. At $N=J$, however, there is 
no need for such correction since the functional derivative with respect to the density at this point is not defined uniquely.
On the other hand, using the XC energy functional, the ionization theorem for Coulombic 
systems leads to the well-known expression for the DD of the XC energy functional: $-A-\epsilon\lumo$. 

Eq. (\ref{mformula}) indicates that the approximation
$
\tilde{E}\app\XC[n]=(1-\omega)E\app\XC[n_{J-1}]+\omega E\app\XC[n_J]
$
misses the different KS kinetic energy contributions leading to the piecewise linear 
features of the energy. ({\em Note}: It does hold for the uniform electron gas where the level spacing is negligible. 
The discrete-state densities returned in that case by the minimization of the kinetic energy are 
negligibly different from those returned by $F$ when both are evaluated at the electron-gas density 
$n$, and $N$ is not an integer). Employing the optimized effective potential method,
\citet{KK13} showed that the linear dependency 
on the number of electrons is almost restored using the functional $\tilde{E}\app\XC$. 
%However, they needed to 
%use the optimized effective potential method for non-integer number of electrons. 
With the kinetic energy contributions 
of Eq. (\ref{mformula}), it is {\em completely} restored.

\begin{figure}[h]
\centering
\subfigure[]{
\includegraphics[scale=0.20]{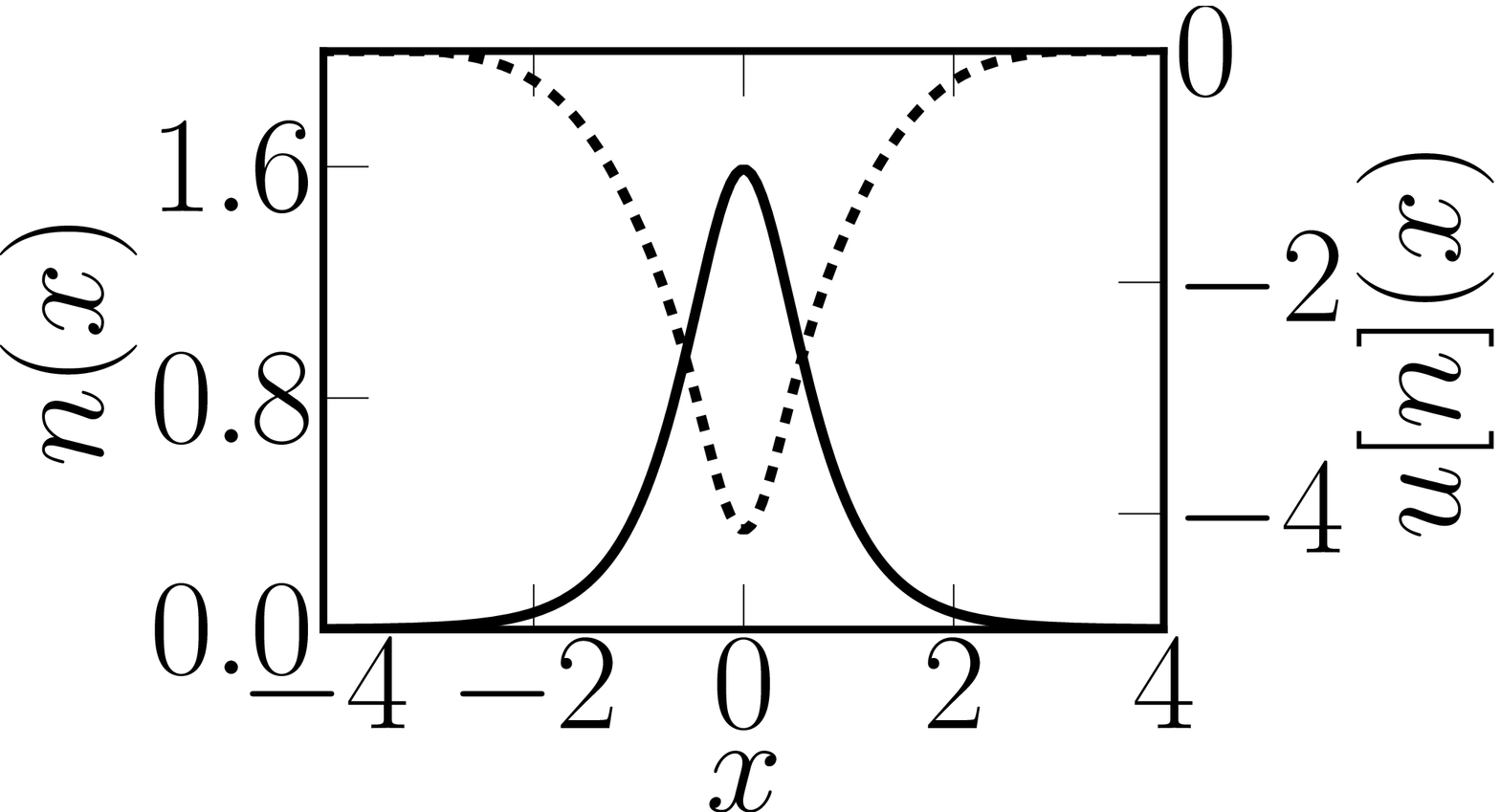}}
\subfigure[]{
\includegraphics[scale=0.17]{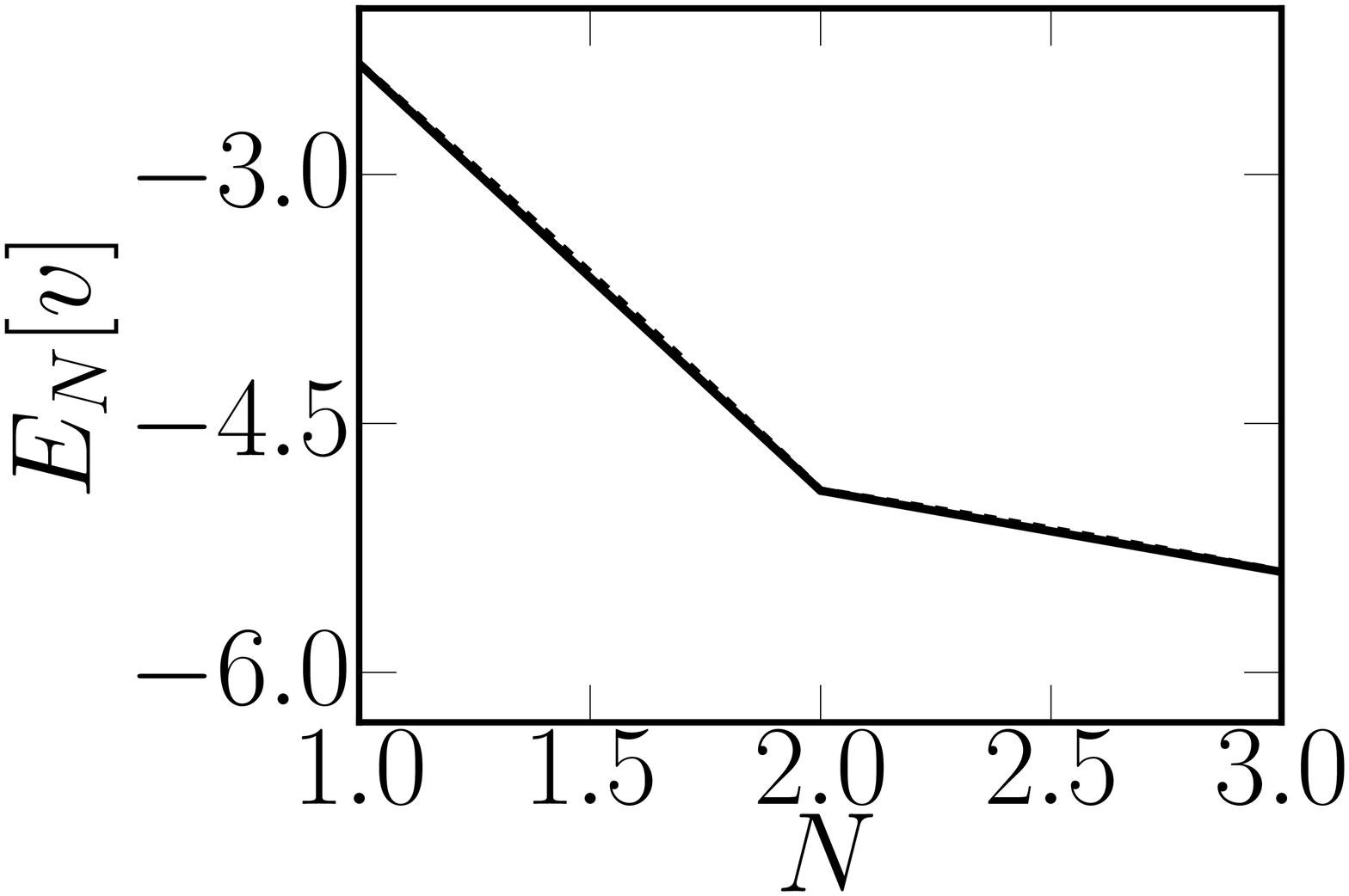}}\\
\subfigure[]{
\includegraphics[scale=0.17]{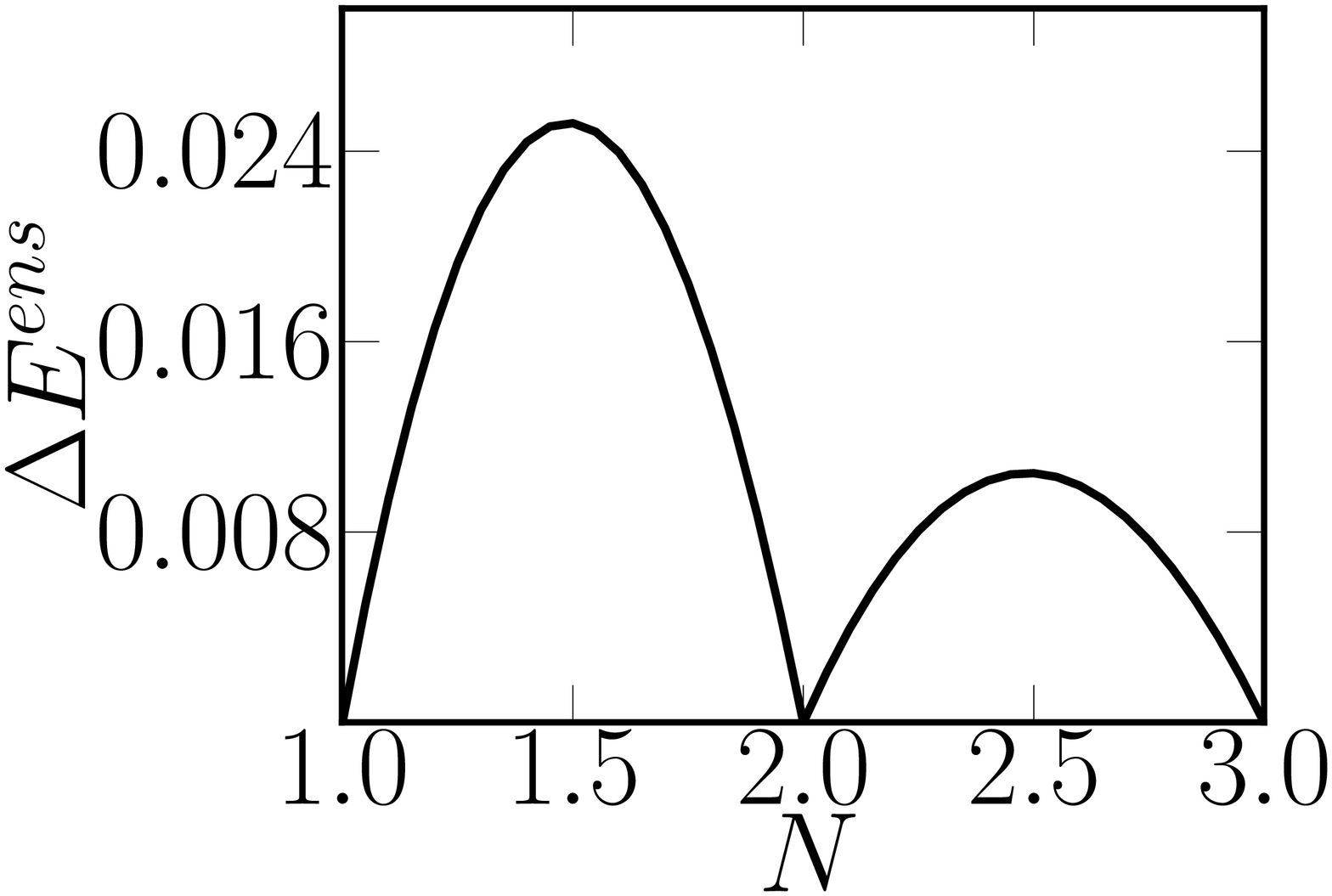}}
\subfigure[]{
\includegraphics[scale=0.17]{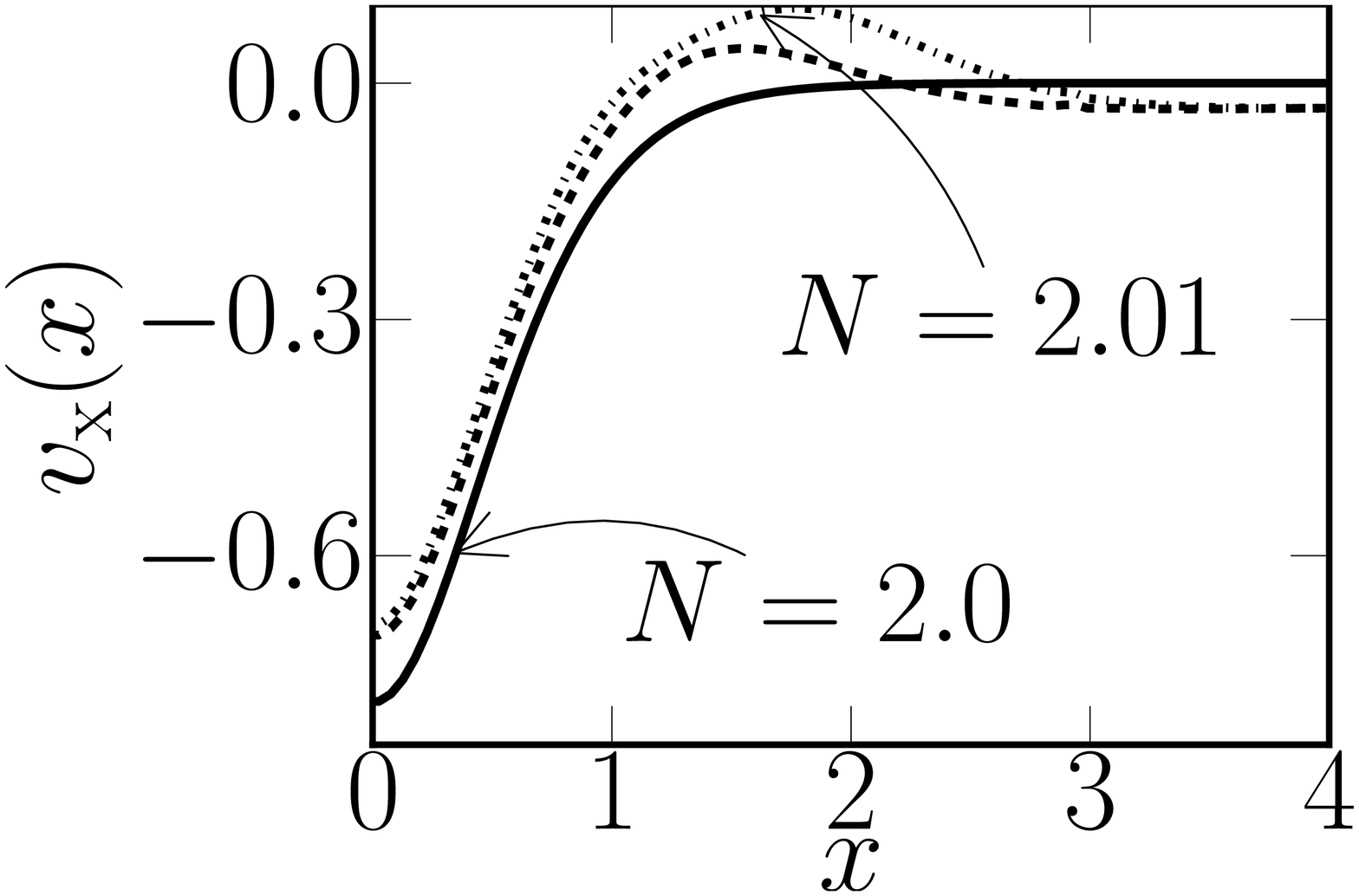}}
\caption{a) Preset ensemble density for $N=2.5$ (solid line), external potential (dashed line). b) Energy as 
a function of $N$ (solid line), approximated energy (dashed line). c) Difference between Eq. (\ref{avenr}) and 
the energy calculated using $E\X[n]=-1/4 \int n^2$ for any $N$. d) Ensemble exchange potential for 
$N=2.0$ (solid line), $N=2.15$ (dashed line), and $N=2.01$ (dashed-dotted line).}
\label{fig1}
\end{figure}

To illustrate our findings, we consider the example of a system of contact-interacting fermions \cite{MB04,RPCP09} described by 
the energy functional $E_v[n_M]=T\s[n_M]+E\H[n_M]+E\X[n_M]+
\int dx v(x)n_M(x)$, where $E\H[n_M]=1/2 \int dx n^2(x)$ and $E\x[n_M]=
-1/4\int dx n_M^2(x)$. 
Suppose that $n_N^{\mr{ref}}(x)=(Na/\pi)\mathrm{sech}(ax)$
is a density of interest with $N=2.5$ and $a=2$. To find the potential 
$u[n]$, we minimize the error functional:
$e^2_N[u]=\lVert \sqrt{n}_N[u]-\sqrt{n}_N^{\mr{ref}}\rVert^2_2$.
The preset density is recovered by solving the KS equations for $N=2$ and $N=3$ and setting 
$n_{2.5}[u](x)=\half n_2[u](x)+\half n_3[u](x)$. Note that the self-consistent procedure has to be applied twice, 
once for $v\x[n_2]=-\half n_2(x)$ and once for $v\x[n_3]=-\half n_3(x)$; in both cases,
the same estimation of $u$ is used. 
We represent $u$ in a spline basis set and $e^2_N[u]$ is minimized with the
Levenberg-Marquardt algorithm \cite{L44,M63}. 
This procedure yields the optimal potential $u[n_{2.5}]$ shown in Fig. \ref{fig1}.a. 
Now we set that potential 
as fixed $v(x)=u(x)$ and calculate the ensemble energy as a function of the number of electrons. 
Fig. \ref{fig1}.b shows the results. The solid line represents the piecewise ensemble interpolation and the dashed 
lines result from setting $E\x[n_N]=-1/4 \int dx n_N^2(x)$ (which lacks the DD) to calculate the energy for any number 
of electrons. The solid and dashed lines look to the eye very close to each other, but their differences are made clear in Fig. \ref{fig1}.c.
This difference is small for the functional chosen. The deviation is more severe for the 3D LDA functional  \cite{CMY08}. 

In Fig. 1d we show the estimation of the DD that results 
from inverting the KS equations for a non-integer number of electrons 
close to $N=2$. To impose the Janak's theorem we minimize the error functional:
\ben
\tilde{e}^2_N[u\ks]=\lVert \sqrt{n}_N[u\ks]-\sqrt{n}_N^{\mr{ref}}\rVert^2_2+(\epsilon_{{\sss H},N}[u\ks]-\epsilon\homo^{\mr{ref}})^2~,
\een
where $n_N^{\mr{ref}}$ is the target ``exact'' ensemble density 
that corresponds to the external potential shown in Fig. \ref{fig1}.a and electron number $N$. 
$\epsilon\homo^{\mr{ref}}$ is the HOMO eigenvalue of the system with $N=3$, obtained from solving 
the KS eqs. with $v\X=-1/2n_3$ and external potential $v$. 
$\tilde{e}^2_N$ is minimized using the conjugate-gradient method \cite{PR69}.  
Because the ionization theorem is not satisfied, the 
potential satisfying $v\X\rightarrow 0$ as $x\rightarrow \pm \infty$
must be shifted by the constant $-A-\epsilon\homo(N=3)$.
In accordance with Eq. (\ref{dd}), the ensemble exchange potential displays its corresponding 
derivative discontinuity. In Fig. 1d, the difference between the curves for $N=2.01$ and $N=2.0$ is $-A[v]-\epsilon\lumo(N=2.0)$. If we shifted the solid curve by $-I[v]-\epsilon\homo(N=2.0)$ ($I[v]=E_1[v]-E_2[v]$) 
and compared the shifted curve (which is $\lim_{N\rightarrow 2^-}v\X$) with the curve for $N=2.01$, we would
observe the discontinuity shown in Eq. (\ref{dd}) around the center of the 1d atom. 
On the other hand, the KS potential far from the center is given by $u\ks(x)\rightarrow \mr{Const.}+
1/(2\phi_2(x))d^2\phi_2/dx^2$. When the number of electrons is slightly 
increased above $N=2$, we are adding a density $\delta n=\epsilon n_3$ with a slower asymptotic decay than 
that of the system with $2$ electrons, causing the discontinuity in Fig. 1.d because $\delta n(x)$ 
only affects the potential at distances that are far from the center. 

\begin{figure}[h]
\centering
\subfigure[]{
\includegraphics[scale=0.18]{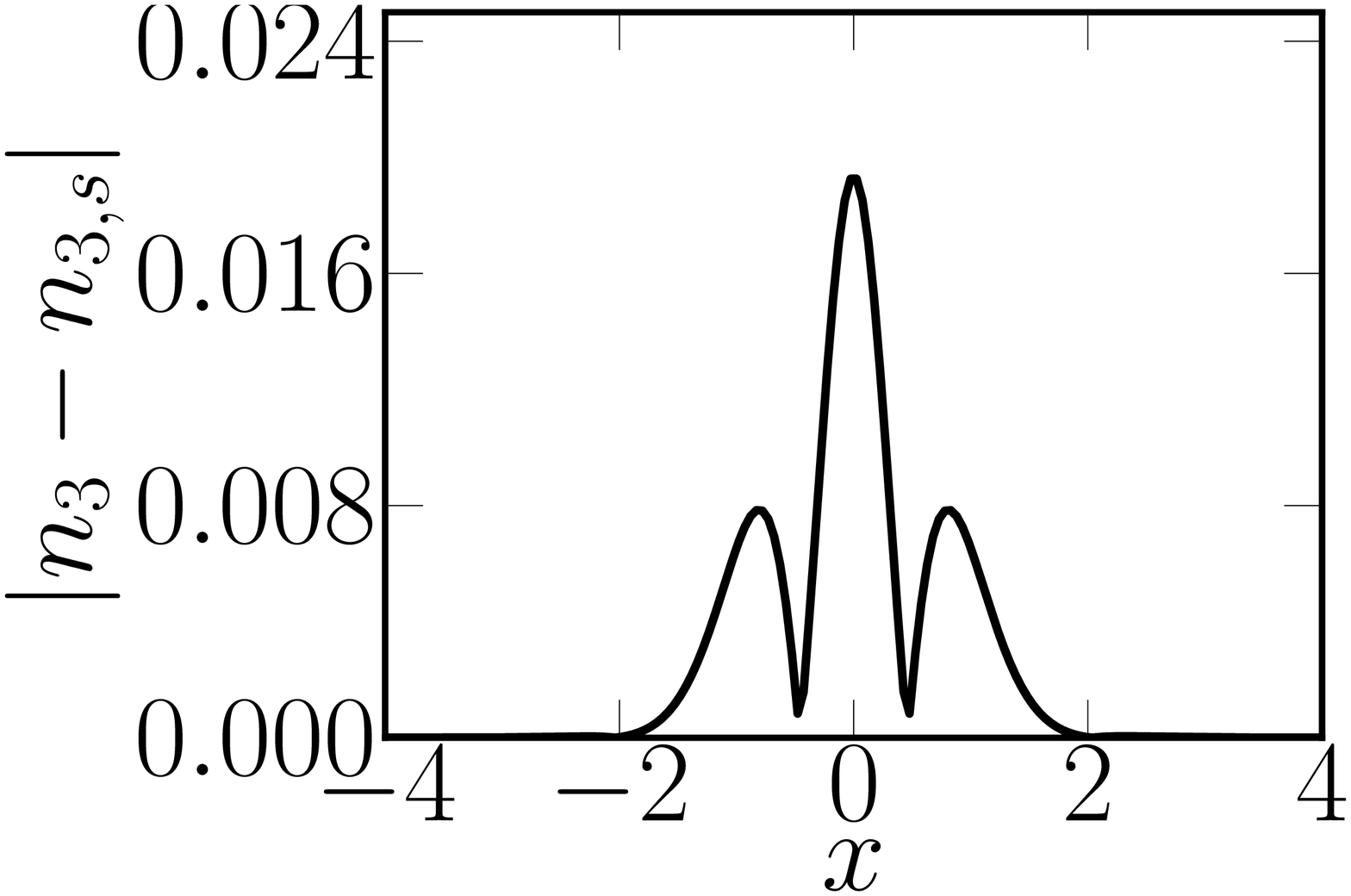}}
\subfigure[]{
\includegraphics[scale=0.18]{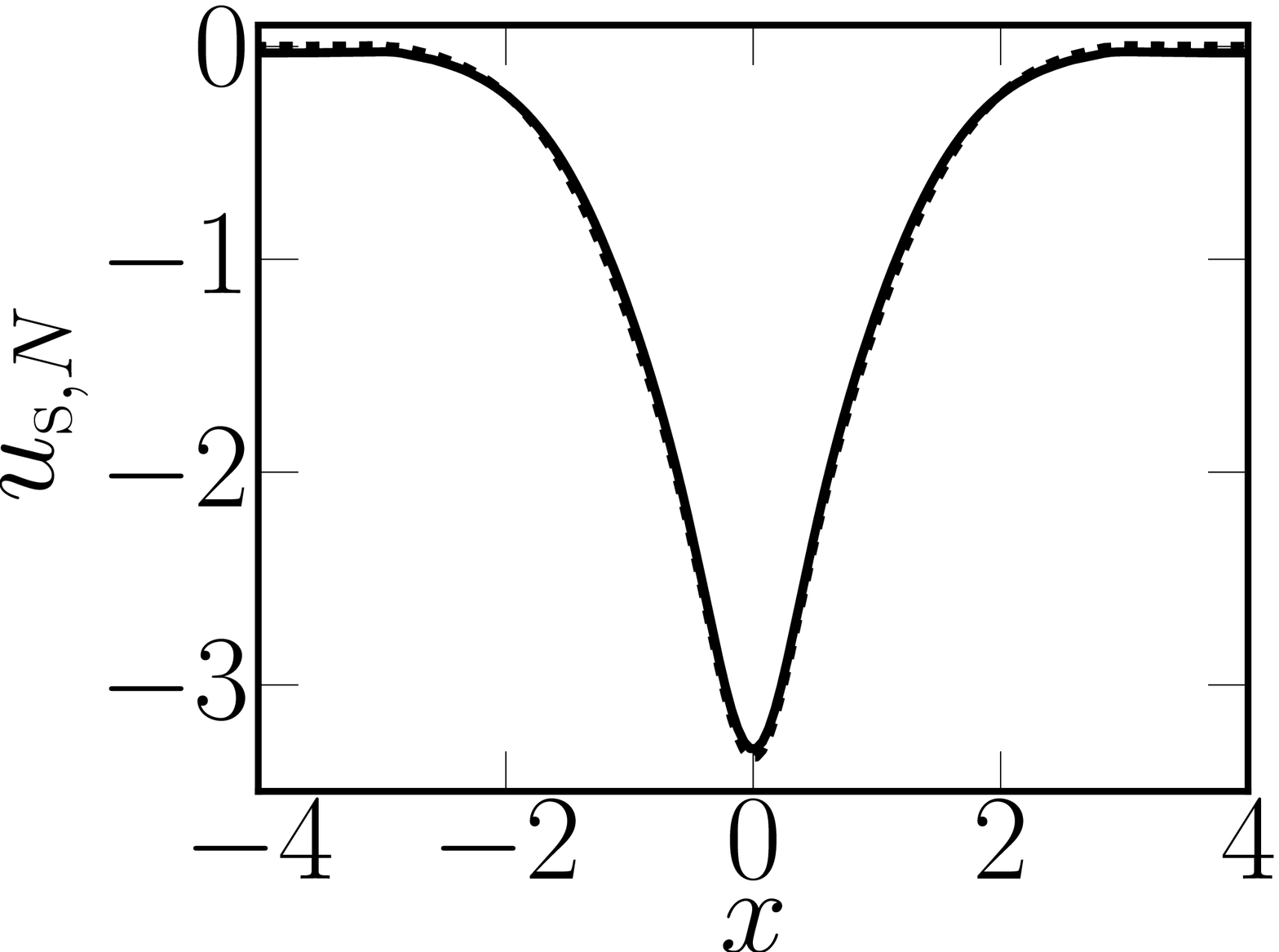}}
\caption{a) Difference between $n_{M}$ and $n_{{\sss s},M}$ for $M=3$; these densities are required 
to yield the density in Fig. 1.a. b) Kohn-Sham potentials corresponding to $N=2.5$ (solid line) and
$N=2.0$ (dashed line).}
\label{fig2}
\end{figure}

We stress that a functional approximation for discrete states is enough to determine, through Eq. (\ref{mformula}), 
an approximation to the XC functional that is also applicable to ensembles. However,
solving the linearity problem in DFA's is not enough to solve the problem of molecular dissociation, 
which is caused by incorrect electron delocalization. A possible solution is to induce 
localization by partitioning a molecule into subsystems or a system-bath complex \cite{CW07}.
In such case, a functional with the correct DD is required since the theory of ensembles 
provides a rigorous framework for defining energy functionals of open systems.
This idea follows the main argument of Ref. \cite{PPLB82} pointing 
to the importance of the XC DD.
%, which was the case of the adiabatic electron transfer between two different 
%neutral atoms separated at certain large distance.

In conclusion, we presented a formal framework to extend density functional approximations of pure-state systems 
to be applicable to densities that integrate to fractional numbers of electrons. The main result, an exact condition, 
is a recursive formula relating the HXC energy with the KS kinetic energy evaluated at the non-interacting bordering 
densities, and the HXC and KS energies evaluated at the bordering interacting densities.
However, the Hohenberg-Kohn-Mermin theorem expressing the densities $n_M[u]$ 
as functionals of $n(\br)$ does not allow us to express $E\XC[n]$ as an explicit functional 
of $n(\br)$, not even when using explicit functionals of the discrete-electron densities. Thus, the 
ensemble $v\XC(\br)$ must be accessed through inversion.

{\em Acknowledgements:} This work was supported by the Office of Basics Energy Sciences, U.S. Department of Energy, 
under grant No. DE-FG02-10ER16196. AW also acknowledges support from an Alfred P. Sloan Foundation Research Fellowship.

\bibliography{dd_refs}

\end{document}